\documentclass[aps,pra,amsfonts,amssymb,amsmath,showpacs,
floatfix,nofootinbib,groupedaddress,superscriptaddress,citesort,twocolumn]{revtex4}

\usepackage{amssymb,amsmath,amsthm}

\theoremstyle{definition}

\theoremstyle{remark}

\usepackage{subfigure}
\usepackage{mathrsfs}
\usepackage{longtable}
\usepackage{amsfonts}
\usepackage{amstext}
\usepackage[usenames]{xcolor}
\usepackage[dvips]{graphicx}
\def\qed{\leavevmode\unskip\penalty9999 \hbox{}\nobreak\hfill
     \quad\hbox{\leavevmode  \hbox to.77778em{%
              \hfil\vrule   \vbox to.675em%
               {\hrule width.6em\vfil\hrule}\vrule\hfil}}
     \par\vskip3pt}

\begin{document}
\title{Complete Characterization of Qubit Masking}

\author{Xiao-Bin Liang}
\email{liangxiaobin2004@126.com.}
\affiliation{School of Mathematics and Computer science, Shangrao Normal University, Shangrao 334001, China}
\affiliation{Quantum Information Research Center, Shangrao Normal University, Shangrao 334001, China}
\author{Bo Li}
\email{libobeijing2008@163.com.}
\affiliation{School of Mathematics and Computer science, Shangrao Normal University, Shangrao 334001, China}
\affiliation{Quantum Information Research Center, Shangrao Normal University, Shangrao 334001, China}
\author{Shao-Ming Fei}
\email{feishm@cnu.edu.cn}
\affiliation{School of Mathematical Sciences, Capital Normal University, Beijing 100048, China}
\affiliation{Max-Planck-Institute for Mathematics in the Sciences, 04103 Leipzig, Germany}

\begin{abstract}
We study the problem of information masking through nonzero linear operators that distribute information encoded in single qubits to the correlations between two qubits.
It is shown that a nonzero linear operator can not mask any nonzero measure set of qubit states. We prove that the maximal maskable set of states on the Bloch sphere with respect to any masker is the ones on a spherical circle. Any states on a spherical circle on the Bloch sphere are maskable, which also proves the conjecture on maskable qubit states given in [Phys. Rev. Lett. 120, 230501 (2018)].
Moreover, we provide explicitly operational unitary maskers for all maskable sets.
As applications, new protocols for secret sharing are introduced.
\end{abstract}

\maketitle

\emph{Introduction.} Due to the properties of linearity (unitarity) of the evolution of a closed quantum system in quantum mechanics,
it is well known that there are several no-go theorems such as the no-cloning theorem \cite{1,2,3}, the no broadcasting theorem and the no-deleting theorem \cite{4,5,6,7}.
Recently, Kavan Modi et. al. \cite{8} considered the problem of quantum information masking based on unitary operators, and obtained the so-called no-masking theorem:
it is impossible to mask all arbitrary qubit states by the same unitary operator.
Different from the decoherence of open systems due to interactions between the system and the environment \cite{9,10,11,12}, the quantum masking means that
the information in subsystems are transferred into the correlations of bipartite systems by unitary operations, such that the final reduced states of any subsystems are identical. Namely, the
subsystems themselves contain no longer the initial information.
No-masking theorem is also different from other no-go theorems such that non-orthogonal states cannot be perfectly cloned or deleted.
In fact, there are many sets containing infinitely many nonorthogonal quantum states which can be masked \cite{8}.

No-go theories are of great significance in information processing like key distribution \cite{Gisin} and quantum teleportation \cite{Bennett,Bouwmeester}, which also results in
studies on such as deterministic or probabilistic cloning \cite{18,19,20}, deleting and purification \cite{21,22,23}.
Hiding information of subsystems into the quantum correlation of composite quantum systems has potential applications in secret sharing \cite{24,M} and quantum cryptography \cite{25}.
Besides some interesting results about the structure of the maskable states, a conjecture has been proposed in \cite{8}: the maskable states corresponding to any masker belong to belong to some
spherical circle on the Bloch sphere.

In this letter, we systematically investigate the masking problem of qubit systems. By showing several theorems we give a complete
characterization of the maskable sets, which also proves the conjecture raised in \cite{8}.
We conclude that the maximal maskable set of states on the Bloch sphere are the ones on a spherical circle.
 All the states on an arbitrary spherical circle on the Bloch sphere are maskable. For each maskable set, we construct an operational masker
by giving an explicit unitary operator. In addition, our results also apply to pseudo-Hermitian ${\cal PT}$-symmetric quantum mechanical systems \cite{13,14,15,16}, where the evolution of a system could be not
unitarian.

\emph{Linear operator and measure of qubit states.}
Let $\mathcal{H}_X$ denote the two dimensional Hilbert space associated with the system $X$.
We say that a linear operator $\mathcal{U}$ masks the quantum
information contained in the set of qubit states, $\{|a_s\rangle_A \in \Omega\subseteq \mathcal{H}_A\}$, if it maps $|a_s\rangle_A$ to $\{|\Psi_s\rangle_{AB} \in \mathcal{H}_A\otimes \mathcal{H}_B\}$ such that all the marginal states of $|\Psi_s\rangle_{AB}$ are identical:
$\rho_A=\mathrm{Tr}_B(|\Psi_s\rangle_{AB}\langle\Psi_s|)$ and $\rho_B=\mathrm{Tr}_A(|\Psi_s\rangle_{AB}\langle\Psi_s|)$ for all $s$. Namely, the reduced states
$\rho_A$ and $\rho_B$ contain no information about the value of $s$. $\Omega$ is said to be the maskable set corresponding to the masker $\mathcal{U}$.

An arbitrary pure qubit state $|p\rangle$ can be written as $|p\rangle=\cos\frac{x}{2}|0\rangle+e^{iy}\sin\frac{x}{2}|1\rangle\equiv|(x,y)\rangle$,
where $x\in [0,\pi]$ and $y\in[0,2\pi)$.
From the domain of the parameters $x$ and $y$, we can define an ``area" measure for a set of qubit states.
The total area of all the qubit states is $\pi\times 2\pi=2\pi^2$, i.e., area measure of the point set $[0,\pi]\times[0,2\pi]$ in the two dimensional plane.
Let $\mathcal{U}$ be a linear operator. For $|p_0\rangle,~|p\rangle\in \mathcal{H}_A$ and $|\Phi_0\rangle,~ |\Phi\rangle \in\mathcal{H}_A\otimes\mathcal{H}_B$ such that
$\mathcal{U}:|p_0\rangle\rightarrow|\Phi_0\rangle$ and $|p\rangle\rightarrow|\Phi\rangle$, we denote
\begin{eqnarray}
\Omega_{\mathcal{U}}(|p_0\rangle)=\{|p\rangle:\mathrm{Tr}_A|\Phi\rangle\langle\Phi|
=\mathrm{Tr}_A|\Phi_0\rangle\langle\Phi_0|,\nonumber\\ ~\mathrm{and}~
\mathrm{Tr}_B|\Phi\rangle\langle\Phi|
=\mathrm{Tr}_B|\Phi_0\rangle \langle \Phi_0|\}.
\end{eqnarray}
We say the set $\Omega_{\mathcal{U}}(|(p_0)\rangle)$ is the largest collections of the maskable states with respect to $|p_0\rangle$ and the linear operator $\mathcal{U}$, that is,
the set $\Omega_{\mathcal{U}}(|p_0\rangle)$ is the maskable set with respect to $|p_0\rangle$ and the linear operator $\mathcal{U}$.

For $|p_0\rangle=|(x_0,y_0)\rangle$, the set $\Omega_{\mathcal{U}}(|(x_0,y_0)\rangle)$ can be regarded as a subset of $[0,\pi]\times[0,2\pi)\subseteq \mathbb{R}^2$. We denote $U((x_0,y_0),\delta)=\{|(x,y)\rangle:(x-x_0)^2+(y-y_0)^2<\delta\}$, $(x_0,y_0)\in(0,\pi)\times(0,2\pi)$,  all the qubit states corresponding to points in the neighborhood of $(x_0,y_0)$.
We will show that the area measure of the set of all maskable states is zero.

Without loss of generality, suppose the linear operator $\mathcal{U}$ acts on the base $|0\rangle,|1\rangle$ as follows,
\begin{eqnarray}
& |0\rangle \rightarrow  a_0|00\rangle  + a_1|01\rangle+c_0|10\rangle+c_1|11\rangle=|\Psi_0\rangle,\nonumber\\
& |1\rangle \rightarrow  b_0|00\rangle  + b_1|01\rangle+d_0|10\rangle+d_1|11\rangle=|\Psi_1\rangle,
\end{eqnarray}
where $a_0,a_1,b_0,b_1,c_0,c_1,d_0,d_1 \in\mathbb{C}$.
For convenience, we denote $|\mu_0\rangle=a_0|0\rangle  + a_1|1\rangle$, $|\mu_1\rangle =c_0|0\rangle+c_1|1\rangle$, $|\nu_0\rangle=b_0|0\rangle  + b_1|1\rangle$ and $|\nu_1\rangle =d_0|0\rangle+d_1|1\rangle.$ Then the map of $\mathcal{U}$ can be rewritten as:
$|0\rangle \rightarrow |0\rangle\otimes|\mu_0\rangle  + |1\rangle\otimes|\mu_1\rangle=|\Psi_0\rangle$,
$|1\rangle \rightarrow  |0\rangle\otimes|\nu_0\rangle  + |1\rangle\otimes|\nu_1\rangle=|\Psi_1\rangle$.
For an arbitrary qubit state, $|(x,y)\rangle=\cos\frac{x}{2}|0\rangle+e^{iy}\sin\frac{x}{2}|1\rangle$ we have, $|\Psi\rangle=\mathcal{U}|(x,y)\rangle=\cos\frac{x}{2}|\Psi_0\rangle +e^{iy}\sin\frac{x}{2}|\Psi_1\rangle.$ The reduced density matrix $\rho_A=\mathrm{Tr}_{B}|\Psi\rangle\langle\Psi|$ is given by
$$
\begin{array}{rcl}
\rho_A&=&f_{00}(x,y)|0\rangle\langle0|+f_{01}(x,y)|0\rangle\langle1|\\[1mm]
&&+f_{10}(x,y)|1\rangle\langle0|+f_{11}(x,y)|1\rangle\langle1|,
\end{array}
$$
where
\begin{equation}\label{fkj}
\begin{array}{ll}
f_{00}(x,y)=&\cos^2(\frac{x}{2})\langle\mu_0|\mu_0\rangle+ \sin^2(\frac{x}{2})\langle\nu_0|\nu_0\rangle\\[1mm]
&+ \Re(\sin(x)e^{-iy}\langle\nu_0|\mu_0\rangle),\\[1mm]
f_{11}(x,y)=&\cos^2(\frac{x}{2})\langle\mu_1|\mu_1\rangle+ \sin^2(\frac{x}{2})\langle\nu_1|\nu_1\rangle\\[1mm]
&+ \Re(\sin(x)e^{-iy}\langle\nu_1|\mu_1\rangle),\\[3mm]
f_{01}(x,y)=&\cos^2(\frac{x}{2})\langle\mu_1|\mu_0\rangle+\sin^2(\frac{x}{2})\langle\nu_1|\nu_0\rangle\\[1mm]
&+\displaystyle\frac{1}{2} \sin(x) (e^{-iy}\langle\nu_1|\mu_0\rangle+e^{iy}\langle\mu_1|\nu_0\rangle), \\[3mm]
f_{10}(x,y)=&\cos^2(\frac{x}{2})\langle\mu_0|\mu_1\rangle+\sin^2(\frac{x}{2})\langle\nu_0|\nu_1\rangle\\[1mm]
&+\displaystyle\frac{1}{2}\sin(x) (e^{-iy}\langle\nu_0|\mu_1\rangle+e^{iy}\langle\mu_0|\nu_1\rangle),
\end{array}
\end{equation}
where $\Re(\cdot)$  stands for the real part.

We first give the following theorem, see proof in section I of Supplementary Material:

\emph{Theorem 1.} For an arbitrary qubit state $|(x_0,y_0)\rangle$, $(x_0,y_0)\in(0,\pi)\times(0,2\pi)$, and arbitrary $\delta>0$, one has $\Omega_{\mathcal{U}}(|(x_0,y_0)\rangle)\nsupseteq U((x_0,y_0),\delta)$, i.e., the neighborhood \\states $U((x_0,y_0),\delta)$ can not be masked by any (non-zero) linear operator $\mathcal{U}$.

In quantum mechanics, the evolution of a closed system is described by unitary operators.
In [8] it has been shown that no unitary masker $\mathcal{U}$ can mask all the qubit states.
Here, from our Theorem 1, we can conclude that

\emph{Corollary 1.} No linear masker $\mathcal{U}$ can mask all the qubit states.

In \cite{13}, generalizing the conventional Hermitian
quantum mechanics, Bender and his colleagues established
the ${\cal PT}$(parity-time)-symmetric quantum mechanics.
In such pseudo-Hermitian quantum mechanical systems,
the Hamiltonians are no longer necessarily Hermitian, but may still
have real eigenvalues \cite{14,15,16}. Moreover, the evolution of such systems is no longer unitary in general. Recently, despite the original motivation to build a new framework
of quantum theory, researchers are also aware of the
importance of simulating the ${\cal PT}$-symmetric systems with
conventional quantum mechanics \cite{hmy}. Our Corollary 1 shows that even (non-unitary) linear operators cannot mask qubit states, namely, it is impossible to mask all the qubit states in ${\cal PT}$-symmetric quantum mechanical systems as long as the evolution is linear.

Furthermore, that the $f_{kl}(x,y)$ in (\ref{fkj}) are constants implies that both the real part $\Re (f_{kl}(x,y))$ and the imaginary $ \Im (f_{kl}(x,y))$ of $f_{kl}(x,y)$ are constant functions.
With respect to $kl=\{00,01,10,11\}$, it means that $f_{kl}(x,y)=c_{kl}$ for some complex constants $c_{kl}$. Denote by $\chi(f)$ either the real part $\Re(f)$ or the imaginary part $\Im(f)$  of $f$. We have the following general form for some complex constants $r_{kl}$,
\begin{eqnarray}\label{15}
\chi(p_{kl}\cos x+q_{kl}\sin x\cos y+h_{kl}\sin x\sin y+r_{kl})=0,
\end{eqnarray}
where the coefficients $p_{kl}$, $q_{kl}$, $h_{kl}$ and $r_{kl}$ are determined by (\ref{fkj}).
For example, from $f_{01}(x,y)=c_{01}$, we have $p_{01}=\frac{\langle\mu_1|\mu_0\rangle-\langle\nu_1|\nu_0\rangle}{2}$,
$q_{01}=\frac{\langle\nu_1|\mu_0\rangle+\langle\mu_1|\nu_0\rangle}{2}$,
$h_{01}=\frac{(\langle\mu_1|\nu_0\rangle-\langle\nu_1|\mu_0\rangle)i}{2}$,
and $r_{01}=(\frac{\langle\mu_1|\mu_0\rangle+\langle\nu_1|\nu_0\rangle}{2}-c_{01})$.
Set $\cos x=Z$, $\sin x\cos y=X$ and $\sin x\sin y=Y$. One has $X^2+Y^2+Z^2=1$. $(X,Y,Z)$ is the point on the unit sphere, which just corresponds to the pure state $|(x,y)\rangle=\cos\frac{x}{2}|0\rangle+e^{iy}\sin\frac{x}{2}|1\rangle$ on the Bloch sphere.
By Theorem 1, $p_{kl}$, $q_{kl}$, $h_{kl}$ and $r_{kl}$ cannot all be zero (otherwise one can deduce that $U((x_0,y_0),\delta)$ can be masked). Hence (\ref{15}) can be viewed as some local plane equations, $\chi(p_{kl}Z+q_{kl}X+ h_{kl}Y+ r_{kl})=0$,
and represents some spherical circles formed by the intersections of some planes and the Bloch sphere.
Unless these spherical circles are in the same plane or essentially the same, their solutions are usually two points or empty set. That is to say, the maskable set corresponding to any masker at maximum is a spherical circle on  the Bloch sphere. Hence, we just need to consider the following conditional function of the maskable set:
\begin{equation}\label{charac}
f(x,y)=p\cos x+q\sin x\cos y+h\sin x\sin y,
\end{equation}
where $p,q,h$ are real coefficients.
 The solution of equations like  $f(x,y)+r=0$,  with real $r$,
at maximum is a curve, as $p,q,h,r$ cannot be all zero. Therefore the area measure of the solutions is zero. We have the following theorem, see proof in section II of Supplementary Material.

\emph{Theorem 2.} No nonzero linear operator can mask a set of qubit states with nonzero Haar measure  on
the Bloch sphere.

\emph{Maskable sets and unitary maskers.}
In the following we present a complete depiction of maskable sets of qubit states.
As a byproduct we answer the disk conjecture
proposed by Kavan Modi et al in [8]: all the maskable states
corresponding to any unitary masker belong to some state points in a disk.
We show that this conjecture is true.

(\ref{charac}) can be converted to\\
$f(x,y)=p\cos x-\sqrt{q^2+h^2}\sin x\cos (y-\theta)$, which is also equivalent to
\begin{eqnarray}
\hbar^\alpha_\theta(x,y)=\cos\alpha\cos x-\sin\alpha\sin x\cos (y-\theta),
\end{eqnarray}
where $\alpha\in[0,\pi)$, $\theta\in[0,2\pi)$, $\cot\alpha=-\frac{p}{\sqrt{q^2+h^2}}$, $\cos\theta=\frac{-q}{\sqrt{q^2+h^2}}$, $\sin\theta=\frac{-h}{\sqrt{q^2+h^2}}$ if $q^2+h^2\neq0$.
If $q^2+h^2=0$, $\alpha=0$.
On the other hand, the states on an arbitrary spherical circle on the Bloch sphere can be expressed as:
\begin{eqnarray}\label{cond}
\mathcal{D}^\alpha_{\theta}(|(x_0,y_0)\rangle)=\{|(x,y)\rangle:
\hbar^\alpha_\theta(x,y)=\hbar^\alpha_\theta(x_0,y_0)\}.
\end{eqnarray}
which corresponds to a circle passing through the point $(x_0,y_0)$ on the Bloch sphere, see
an intuitive description shown in Fig. 1 and Fig. 2.

\begin{figure}[!htbp]
\includegraphics[scale=0.6]{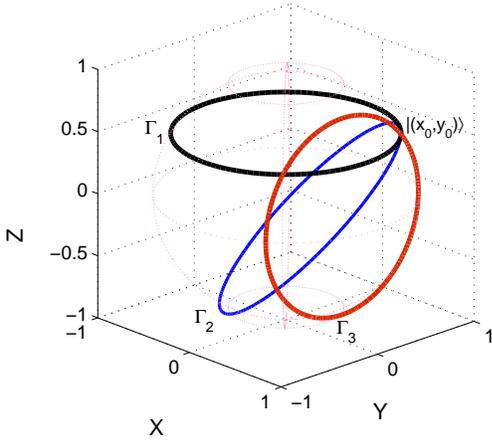}
\caption{Typical maskable sets passing through point $(x_0,y_0)$ (associated with the state $|(x_0,y_0)\rangle$), where $x_0=\frac{\pi}{3}$, $y_0=\frac{\pi}{4}$.
Line $\Gamma_1$ is for $\mathcal{D}^0_{0}(|(x_0,y_0)\rangle)$, which is parallel to the X-Y plane. Counterclockwise, line $\Gamma_2$ is for  $\mathcal{D}^{\pi/4}_{\pi/4}(|(x_0,y_0)\rangle)$, and line $\Gamma_3$ for $\mathcal{D}^{\pi/2}_{0}(|(x_0,y_0)\rangle)$.}\label{fig_1}
\end{figure}

\begin{figure}[!htbp]
\includegraphics[scale=0.7]{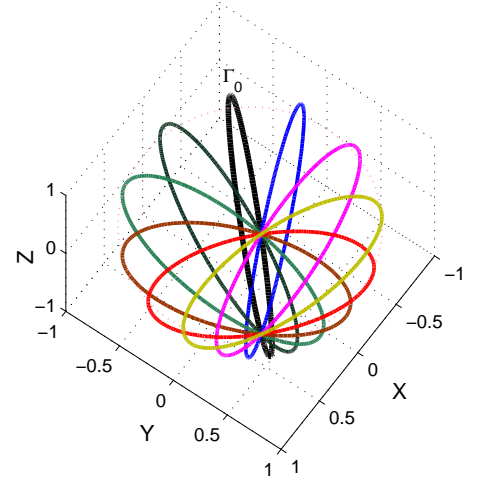}
\caption{Maskable sets passing through the point $(x_0,y_0)$ which are all vertical to the X-Y plane, where $x_0=\frac{\pi}{6}$ and $y_0=\frac{\pi}{4}$. Line $\Gamma_0$ is for $\mathcal{D}^{\pi/2}_{\pi/4}(|(x_0,y_0)\rangle)$, and in counterclockwise, $\mathcal{D}^{\pi/2}_{\pi/4+k\pi/8}(|(x_0,y_0)\rangle)$, $k=0,1,...,7$.}\label{fig_2}
\end{figure}

In the following, we construct an isometry operator to  mask the spherical circle sets
of states. We define a masker $\mathcal{S}_\theta^\alpha$ such that
$$
\mathcal{S}_\theta^\alpha|0\rangle|b\rangle=|0\rangle|u_0\rangle\  + |1\rangle|u_1\rangle,~
\mathcal{S}_\theta^\alpha|1\rangle|b\rangle=|0\rangle|v_0\rangle\  + |1\rangle|v_1\rangle,
$$
where
\begin{equation}
\begin{array}{l}
|u_0\rangle=\displaystyle\frac{\sqrt{2}}{2}(\cos(\frac{\alpha}{2}) e^{(\theta+\pi/4)i}|0\rangle+\cos(\frac{\alpha}{2}) e^{(\theta+\pi/4)i}|1\rangle), \\[1mm]
|u_1\rangle=\displaystyle\frac{\sqrt{2}}{2}(\sin(\frac{\alpha}{2}) e^{(\theta-\pi/4)i}|0\rangle-\sin(\frac{\alpha}{2} ) e^{(\theta-\pi/4)i}|1\rangle), \\[1mm]
|v_0\rangle=\displaystyle -\frac{\sqrt{2}}{2}(\sin(\frac{\alpha}{2}) e^{\pi i/4}|0\rangle+\sin(\frac{\alpha}{2}) e^{\pi i/4}|1\rangle), \\[1mm]
|v_1\rangle=\displaystyle\frac{\sqrt{2}}{2}(\cos(\frac{\alpha}{2}) e^{-\pi i/4}|0\rangle-\cos(\frac{\alpha}{2}) e^{-\pi i/4}|1\rangle).
\end{array}
\end{equation}
It is easily verified that $\mathcal{S}_\theta^\alpha$ is an isometry masker which can be always
realized by a unitary operator on two-qubit space.
We now prove that the states $\mathcal{D}^\alpha_{\theta}(|(x_0,y_0)\rangle)$ can always be masked by $\mathcal{S}_\theta^\alpha$.
That is to say, the states in the set $\mathcal{D}^\alpha_{\theta}(|(x_0,y_0)\rangle)$ will be
mapped to states in $\mathcal{H}_A\otimes\mathcal{H}_B$ by $\mathcal{S}_\theta^\alpha$, such that all their reduced states are identical. The maximal maskable sets of states are the ones on spherical circles on the Bloch sphere.

\emph{Theorem 3.} All the states $\mathcal{D}^\alpha_{\theta}(|(x_0,y_0)\rangle)$ associated with an arbitrary spherical circle passing through the point $(x_0,y_0)$ on the Bloch sphere can be masked by $\mathcal{S}_\theta^\alpha$.

\emph{Proof.} All the qubit states $|(x,y)\rangle\in\mathcal{D}^\alpha_{\theta}(|(x_0,y_0)\rangle)$
satisfy the condition (\ref{cond}), $\hbar^\alpha_\theta(x,y)=\hbar^\alpha_\theta(x_0,y_0)$.
Denote $|\Psi\rangle= \mathcal{S}_\theta^\alpha |(x,y)\rangle$. The reduced density matrices $\rho_{A,B}=\mathrm{Tr}_{B,A}|\Psi\rangle\langle\Psi|$ are given by
$$
\begin{array}{l}
\rho_A= (\frac{1}{2}+\frac{1}{2}\hbar^\alpha_\theta(x,y))|0\rangle\langle0|
+(\frac{1}{2}-\frac{1}{2}\hbar^\alpha_\theta(x,y))|1\rangle\langle1|,\\[2mm]
\rho_B=\frac{1}{2}|0\rangle\langle0|+\frac{1}{2}|1\rangle\langle1|+
\frac{1}{2}\hbar^\alpha_\theta(x,y)(|0\rangle\langle1|+|1\rangle\langle0|).
\end{array}
$$
According to the condition that $\hbar^\alpha_\theta(x,y)$ is constant, we get that $\rho_A$ and $\rho_B$ are fixed matrices. Hence, $\Omega_{\mathcal{S}_\theta^\alpha}(|(x_0,y_0)\rangle)\supseteq \mathcal{D}^\alpha_{\theta}(|(x_0,y_0)\rangle)$.
Namely, arbitrary states on a spherical circle on the Bloch sphere can be masked.
$\blacksquare$

Since any three points lie on same sperical circle of  the Bloch sphere, we have the following
conclusion.

\emph{Corollary 2.} Any three different qubit states can be masked by the same masker.

\emph{Remark} We have shown that all the states on an arbitrary spherical circle passing through the point $(x_0,y_0)$ on the Bloch sphere can be masked by the same masker $\mathcal{S}_\theta^\alpha$.
For instance, $\mathcal{D}^0_{0}(|(x_0,y_0)\rangle)$ and $\mathcal{D}^{\pi/2}_{\theta}(|(x_0,y_0)\rangle)$ are maskable states in circles on the Bloch sphere that are parallel and vertical to the X-Y plane, respectively. These maskable states $\mathcal{D}^\alpha_{\theta}(|(x_0,y_0)\rangle)$ are uncountably
infinitely many.
Our masker $\mathcal{S}_\theta^\alpha$ for qubit case works for arbitrary states.
Such masker is not necessarily unique for specific maskable sets. For example,  $\mathcal{D}^0_{0}(|(x_0,y_0)\rangle)$ can be masked by either $\mathcal{S}_0^0$ or
$\mathcal{S}^\sharp$ given in \cite{8}. Nevertheless, here besides just a proof of the existence of masker,
we also present a uniform constructive and operational way of masking, which can be practically used in quantum information processing such as secret sharing and quantum cryptography.

\emph{Applications of the maskers $\mathcal{S}_\theta^\alpha$.}
The maskable sets can be used for no qubit commitment \cite{8} and quantum secret sharing \cite{Zhen,27,28} etc..
Here we introduce an application to protocols for unlocking secret information under the cooperation of certain observables.
Alice encodes the message $(x_0,y_0)$ into the state $|(x_0,y_0)\rangle$.
By applying a set of maskers, $\mathcal{S}_{\theta_k}^{\alpha_k}$, $k=1,...,N$, she gets a set of qubit pairs $A$ and $B$ in states $|\Psi_k\rangle_{AB}$.
Alice keeps the qubits $A$s, and send the qubits $B$s to \{$Bob_1, Bob_2,..., Bob_N$\}, respectively.
The Bobs can only obtain information about the reduced states, and cannot decode the information by local quantum operations without classical communication,
even if Alice informed them of the maskers $\mathcal{S}_{\theta_k}^{\alpha_k}$.
$Bob_k$ only knows that the message must be one of the $(x,y)$ in the set of maskable states
$\mathcal{D}^{\alpha_k}_{\theta_k}(|(x_0,y_0)\rangle)=\{|(x,y)\rangle:\, \hbar^{\alpha_k}_{\theta_k}(x,y)=\hbar^{\alpha_k}_{\theta_k}(x_0,y_0)\}$, namely, one of the points on the spherical circle
with respect to the masker $\mathcal{S}_{\theta_k}^{\alpha_k}$.
However, if some Bobs cooperate together, they generally can obtain the encoded message $(x_0,y_0)$.

For example, if Alice uses maskers $\mathcal{S}_{0}^{\alpha_k}$, $\alpha_k=k\pi/n$, $k=1,2,...,n-1$, $n\geq 3$, then any two Bobs
cooperate together, they can obtain the encoded message $(0,0)$, since two different spherical circles
have only one unique intersecting point $(0,0)$, see Fig. 3.

\begin{figure}[!htbp]
\includegraphics[scale=0.6]{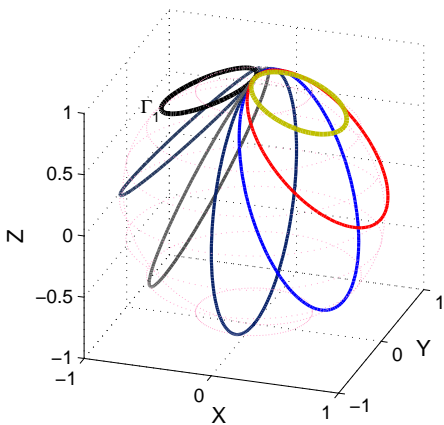}
\caption{Maskable sets passing through point $(0,0)$, the line $\Gamma_1$ is for $\mathcal{D}^{\pi/8}_{0}(|(0,0)\rangle)$, and in counterclockwise, $\mathcal{D}^{k\pi/8}_{0}(|(0,0)\rangle)$, $k=1,2,...7$.}\label{fig_3}
\end{figure}

If Alice uses maskers $\mathcal{S}_{\theta_k}^{\alpha_k}$, $\theta_k,\alpha_k=k\pi/n$, $k=1,2,...,n-1$, $n\geq 4$, then any three Bobs cooperating together can obtain the message $(x_0,y_0)$, since in this case, any three spherical circles (their respective planes) are not parallel to the same straight line, they have only one unique intersecting point $(x_0,y_0)$, see Fig. 1, although
any two spherical circles have two intersecting points.
For instance, Alice may use the masker $\mathcal{S}_0^{0}$, $\mathcal{S}_0^{\pi/2}$ and $\mathcal{S}_{\pi/2}^{\pi/2}$ to mask the qubit state $|(x_0,y_0)\rangle$.
From (\ref{cond}), what $Bob_1$, $Bob_2$ and $Bob_3$ know are some $(x,y)$ satisfying $\cos x=\cos x_0$, $\sin x\,\cos y=\sin x_0\,\cos y_0$
and $\sin x\,\sin y=\sin x_0\,\sin y_0$, respectively. Hence, they can decode the message by classical communications.
Nevertheless, if Alice uses maskers $\mathcal{S}_{\theta_k}^{\pi/2}$, $\theta_k\in[0,2\pi)$, then the message $(x_0,y_0)$ can never be decoded,
despite of the number of Bobs cooperating together, since in this case all spherical circles have the same two intersecting points, see Fig. 2.
As the maskers $\mathcal{S}_{\theta}^{\alpha}$ are infinitely many, the message may be distributed to arbitrary many receivers.
This protocol is different from the one in which only one masker is used to mask many states in the maskable set for secret sharing.

\emph{Conclusion.} In summary, we have presented a complete characterization of the problem of qubit masking. We have shown that
nonzero linear operators can not mask nonzero measure set of qubit states.
As in the proof we used general linear operators instead of unitary operators,
our conclusions also apply to pseudo-Hermitian ${\cal PT}$-symmetric quantum mechanical sys-tems for non-unitarian evolutions \cite{13,14,15,16}.
Hence, it is also impossible to mask all the qubit states in ${\cal PT}$-symmetric quantum mechanics.
Moreover, it has been demonstrated that
the maximum maskable sets of states on the Bloch sphere are on spherical circles, and the states on an arbitrary spherical circle are maskable.
As a byproduct, we proved the ``disk conjecture" raised in \cite{8}. Most of all, we have provided a unified form of operational maskers $\mathcal{S}_\theta^\alpha$ for each maskable set,
which may be of great use in practice in applications such as secret sharing, quantum cryptography
and future quantum communication protocols. Our results may also highlight further studies on masking hight dimensional states.

\bigskip
\noindent {\bf Acknowledgments}
This work is supported by NSFC under Nos 11765016 and 11675113, Jiangxi Education Department Fund (KJLD14088), and Beijing Municipal Commission of Education (KZ201810028042).
Xiao-Bin Liang and Bo Li contribute equally to this work.

\section{Supplemental Material}

\emph{Proof of Theorem 1-}
Now suppose $\mathcal{U}$ can mask  the neighborhood states ${U}((x_0,y_0),\delta)$. Then for an arbitrary qubit state $|(x,y)\rangle\in {U}((x_0,y_0),\delta)$, the reduced state $\rho_A$ should not depend on $x$ and $y$,
$f_{kl}(x,y)$ given by (3) in the main text are constant functions on the set ${U}((x_0,y_0),\delta)$.
Namely, $\Re(f_{kl}(x,y))$ and $\Im(f_{kl}(x,y))$ are also constant. Thus $\forall (x,y)\in U((x_0,y_0),\delta)$, the partial derivatives with respect to $x$ and $y$ must be zero,
$$
\begin{array}{l}
\displaystyle\frac{\partial\Re(f_{00}(x,y))}{\partial y}=0,~~~\frac{\partial\Re(f_{11}(x,y))}{\partial y}=0,\\
\displaystyle\frac{\partial\Re(f_{01}(x,y))}{\partial y}=0,~~~\frac{\partial\Im(f_{01}(x,y))}{\partial y}=0.
\end{array}\eqno(S1)
$$

First consider $f_{00}(x,y)$. Denote $\langle\nu_0|\mu_0\rangle=a+ib$, where $a$ and $b$ are real numbers. Then $\Re(\sin (x)e^{-iy}\langle\nu_0|\mu_0\rangle)=\sin (x)( a\cos y+b\sin y)$, and $\frac{\partial \Re f_{00}}{\partial y}=\sin (x)(a\cos y-b\sin y)\equiv0$, $\forall (x,y)\in {U}((x_0,y_0),\delta)$, requires that $a=0$ and $b=0$. Hence we obtain  $\langle\nu_0|\mu_0\rangle=0$.
In a similar way, from $\frac{\partial \Re f_{11}}{\partial y}=0$ we have $\langle\nu_1|\mu_1\rangle=0$.

Now we consider the partial derivative of $f_{01}(x,y)$. Denote $\langle\nu_1|\mu_0\rangle=c+id$ and $\langle\mu_1|\nu_0\rangle=s+it$, where $c, d, s, t$ are all real numbers. Then
\begin{eqnarray}\nonumber
\Re(\sin(x)e^{-iy}\langle\nu_1|\mu_0\rangle/2+\sin(x)e^{iy}\langle\mu_1|\nu_0\rangle/2)
=\nonumber\\ \frac{1}{2} \sin(x) ((s+c)\cos y+(d-t)\sin y),\nonumber \\
\Im(\sin(x)e^{-iy}\langle\nu_1|\mu_0\rangle/2+\sin(x)e^{iy}\langle\mu_1|\nu_0\rangle/2)
=\nonumber\\ \frac{1}{2}\sin(x) ((d+t)\cos y+(s-c)\sin y).\nonumber
\end{eqnarray}
By $\frac{\partial \Re f_{01}}{\partial y}=\frac{\partial \Im f_{01}}{\partial y}=0$ for all $(x,y)\in {U}((x_0,y_0),\delta)$, one gets that
$$
\begin{array}{l}
\sin (x) (-(s+c)\sin y+(d-t)\cos y)\equiv 0,\\
\sin( x) (-(d+t)\sin y+(s-c)\cos y)\equiv 0.
\end{array}\eqno(S2)
$$
Because $\sin x$, $\sin y$ and $\cos y$ are not identically zero $\forall (x,y)\in {U}((x_0,y_0),\delta)$,
we assert that $c=d=s=t=0$. Otherwise, the solution of (S2) is at most a curve of elementary function, and the area measure of all $(x, y)$ satisfying (S2) must be zero, which contradicts to the maskable assumptions. Namely, we have $\langle\nu_1|\mu_0\rangle=0$ and $\langle\nu_0|\mu_1\rangle=0$.

Altogether, from (S1) one obtains
$$
\begin{array}{l}
\langle\nu_0|\mu_0\rangle=\langle\nu_1|\mu_1\rangle=\langle\nu_1|\mu_0\rangle=\langle\nu_0|\mu_1\rangle=0.
\end{array}\eqno(S3)
$$
Substituting (S3) into (3) in the main text we have
$\forall (x,y)\in {U}((x_0,y_0),\delta)$,
\begin{eqnarray}
\frac{\partial f_{00}(x,y)}{\partial x}=-\frac{1}{2}\langle\mu_0|\mu_0\rangle\sin x +\frac{1}{2}\langle\nu_0|\nu_0\rangle\sin x\equiv0, \nonumber\\~~~
\frac{\partial f_{11}(x,y)}{\partial x}=-\frac{1}{2}\langle\mu_1|\mu_1\rangle\sin x +\frac{1}{2}\langle\nu_1|\nu_1\rangle\sin x\equiv0, \nonumber\\
\frac{\partial f_{01}(x,y)}{\partial x}=-\frac{1}{2}\langle\mu_1|\mu_0\rangle\sin x+\frac{1}{2}\langle\nu_1|\nu_0\rangle\sin x\equiv0,\nonumber
\end{eqnarray}
which give rise to
$$
\begin{array}{l}
\langle\mu_1|\mu_0\rangle=\langle\nu_1|\nu_0\rangle,~
\langle\mu_i|\mu_i\rangle=\langle\nu_i|\nu_i\rangle,~~i=0,1.
\end{array}\eqno(S4)
$$
Obviously, $|\mu_0\rangle$ and $|\mu_1\rangle$ can not be all zero. Assuming
$|\mu_0\rangle\neq0$, by (S4) one gets $|\nu_0\rangle\neq0$. Since $\langle\nu_0|\mu_1\rangle=\langle\nu_0|\mu_0\rangle=0$, from (S3) on has $|\mu_1\rangle=\lambda_1|\mu_0\rangle$. Similarly, from
$\langle\nu_0|\mu_0\rangle=\langle\nu_1|\mu_0\rangle=0$, one obtains
$|\nu_1\rangle=\lambda_2|\nu_0\rangle$.
At last, we have $\lambda_1=\lambda_2$ due to $\langle\mu_1|\mu_0\rangle=\langle\nu_1|\nu_0\rangle$.
Namely, there exists $\lambda$ such that $|\mu_1\rangle=\lambda|\mu_0\rangle$ and $|\nu_1\rangle=\lambda|\nu_0\rangle.$ Therefore, the linear operator $\mathcal{U}$ gives rise to
the following map,
$|0\rangle \rightarrow (|0\rangle + \lambda|1\rangle)\otimes|\mu_0\rangle$, $|1\rangle \rightarrow (|0\rangle + \lambda|1\rangle)\otimes|\nu_0\rangle$.

The reduced density matrix $\rho_B$ is then of the form,
$$
\begin{array}{rcl}
\rho_B&=&\displaystyle(1+|\lambda|^2)|(\cos^2\frac{x}{2}|\mu_0\rangle\langle\mu_0| +\sin^2\frac{x}{2}|\nu_0\rangle\langle\nu_0|\\
&&+\displaystyle\frac{1}{2}\sin(x)e^{-iy}|\mu_0\rangle\langle\nu_0|+\frac{1}{2}\sin (x)e^{iy}|\nu_0\rangle\langle\mu_0|),
\end{array}
$$
where $|\mu_0\rangle=a_0|0\rangle +a_1|1\rangle$ and $|\nu_0\rangle=b_0|0\rangle + b_1|1\rangle.$
Repeating the same analysis as $\rho_A$, we can draw conclusions parallel to (S3) and (S4).
Since $\mathcal{U}$ is a nonzero linear
operator, one may assume that $a_0\neq0$. Notice that $a_0b_0^*=0$ and $|a_0|=|b_0|$ cannot be true simultaneously. Therefore, for any $(x_0,y_0)\in (0,\pi)\times(0,2\pi)$,  and its  neighborhood ${U}((x_0,y_0),\delta)$, any  nonzero linear operator $\mathcal{U}$ cannot mask the neighborhood. $\blacksquare$

\emph{Proof of Theorem 2-}
Denote $\mathcal{B}=(0,\pi)\times [0,2\pi)\cup\{(0,0),(\pi,0)\}$.
The following relations
\begin{eqnarray}
\begin{cases}
 |(x,y)\rangle=\cos\frac{x}{2}|0\rangle +e^{iy}\sin\frac{x}{2}|1\rangle, (x,y)\in(0,\pi)\times [0,2\pi);
 \cr |(0,0)\rangle=|0\rangle;~ |(\pi,0)\rangle=|1\rangle\end{cases}
\nonumber
\end{eqnarray}
give a bijection between the planar point set $\mathcal{B}$ and all the states on the Bloch sphere.
Denote $\mathcal{B}_n=[0+\frac{1}{n},\pi-\frac{1}{n}]\times[0+\frac{1}{n},2\pi-\frac{1}{n}]$ which is a bounded closed set. Then
$\mathcal{B}_1\subseteq\mathcal{B}_2\subseteq...\subseteq\mathcal{B}_n\subseteq...\subseteq\mathcal{B}$ and
$\lim_{n\rightarrow+\infty}M[C(\mathcal{B}_n)]=0$, where $M[\cdot]$ is the Lebesgue measure, $C(\mathcal{B}_n)=\mathcal{B}\backslash\mathcal{B}_n$ is the complementary set of $\mathcal{B}_n$.

Now suppose there exists maskable set $A\subset \mathcal{B}$ such that $M[A]>0$,
then $\exists~  n$, such that $M[A\cap\mathcal{B}_n]>0$, since
$$
M[A]=M[A\cap\mathcal{B}_n]+M[A\cap C(\mathcal{B}_n)].
$$

By Theorem 1, and according to the previous results about the conditional function of the maskable set (5) in the main text,  we know that the maskable set in ${U}((x_0,y_0),\delta)$  at the maximum is a curve of elementary function, with its area measure zero. Namely, $\forall~(x_0,y_0)\in \mathcal{B}_n\subset (0,\pi)\times(0,2\pi)$, $\forall~\delta>0$ and ${U}((x_0,y_0),\delta)$, it holds that $M[A\cap{U}((x_0,y_0),\delta)]=0$.
Because
\begin{eqnarray}
\bigcup_{(x_0,y_0)\in \mathcal{B}_n}{U}((x_0,y_0),\delta)\supseteq \mathcal{B}_n, \nonumber \end{eqnarray}
which  is  an open cover of the bounded closed set $\mathcal{B}_n$, by the finite covering theorem, there exists finite subcover
\begin{eqnarray} \bigcup^N_{k=1}{U}((x_k,y_k),\delta)\supseteq \mathcal{B}_n\nonumber\end{eqnarray}
and $M[A\cap{U}((x_k,y_k),\delta)]=0$, for $k=1,2,...,N$.
Therefore,
$$
\begin{array}{l}
M[A\cap\mathcal{B}_n]\leq M[A\cap\bigcup^N_{k=1}{U}((x_k,y_k),\delta)]\\[2mm]
\leq \sum^N_{k=1}M[A\cap{U}((x_k,y_k),\delta)]=0.
\end{array}
$$
This is a contradiction to $M[A\cap\mathcal{B}_n]>0$.

Hence the assumptions that the maskable sets $A\subset \mathcal{B}$ and $M[A]>0$ is wrong. That is to say, $M[A]$ must be zero. Note $\widehat{\mathcal{B}}=(0,\pi)\times(0,2\pi)$, and $\widehat{S(E)}= \{\cos\frac{x}{2}|0\rangle +e^{iy}\sin\frac{x}{2}|1\rangle : 0<x<\pi, 0<y<2\pi\}$,  $\widehat{S(E)}$ is a piont set of on
the Bloch sphere ( the Bloch sphere remove a half semicircular arc).
Due to the bijective mapping from the open set of nonzero Lebesgue  measure on $\widehat{\mathcal{B}}$ to the open set of nonzero Haar measure on the Bloch sphere, the Lebesgue  measure on $\widehat{\mathcal{B}}$ is equivalent to the Haar measure on $\widehat{S(E)}$. Hence, any linear operator can not mask any nonzero Haar measure set on the Bloch sphere.
$\blacksquare$

\end{document}